\documentclass[twocolumn]{aastex63}

\usepackage{graphicx}
\usepackage[T1]{fontenc}
\usepackage{aecompl}
\usepackage[]{amsmath}
\usepackage{color}
\usepackage{xspace}
\usepackage{soul}

\newcommand{\msun}{\mathrm{M}_\odot}
\graphicspath{{figures/}}

\newcommand{\ud}{\mathrm{d}}

\def\lsim{ \lower .75ex \hbox{$\sim$} \llap{\raise .27ex \hbox{$<$}} }

\submitjournal{ApJ}

\shorttitle{}
\shortauthors{Zhang et al.}
\graphicspath{{./}{figures/}}

\begin{document}


\title{Using two point correlation functions to understand the assembly histories of Milky Way-like galaxies}

\correspondingauthor{Wenting Wang}
\email{wenting.wang@sjtu.edu.cn}

\author{Yike Zhang}
\affiliation{Department of Astronomy, Shanghai Jiao Tong University, Shanghai 200240, China}
\affiliation{Shanghai Key Laboratory for Particle Physics and Cosmology, Shanghai 200240, China}
\author[0000-0002-5762-7571]{Wenting Wang}
\affiliation{Department of Astronomy, Shanghai Jiao Tong University, Shanghai 200240, China}
\affiliation{Shanghai Key Laboratory for Particle Physics and Cosmology, Shanghai 200240, China}
\author[0000-0002-8010-6715]{Jiaxin Han}
\affiliation{Department of Astronomy, Shanghai Jiao Tong University, Shanghai 200240, China}
\affiliation{Shanghai Key Laboratory for Particle Physics and Cosmology, Shanghai 200240, China}
\author{Xiaohu Yang}
\affiliation{Department of Astronomy, Shanghai Jiao Tong University, Shanghai 200240, China}
\affiliation{Shanghai Key Laboratory for Particle Physics and Cosmology, Shanghai 200240, China}
\author{Vicente Rodriguez-Gomez}
\affiliation{Instituto de Radioastronom\'ia y Astrof\'isica, Universidad Nacional Aut\'onoma de M\'exico, Apdo. Postal 72-3, 58089 Morelia, Mexico}
\author{Carles G. Palau}
\affiliation{Department of Astronomy, Shanghai Jiao Tong University, Shanghai 200240, China}
\affiliation{Shanghai Key Laboratory for Particle Physics and Cosmology, Shanghai 200240, China}
\author{Zhenlin Tan}
\affiliation{Department of Astronomy, Shanghai Jiao Tong University, Shanghai 200240, China}
\affiliation{Shanghai Key Laboratory for Particle Physics and Cosmology, Shanghai 200240, China}




\begin{abstract}

 The two point correlation function (2PCF) is a powerful statistical tool to measure galaxy clustering. Although 2PCF has also been used to study the clustering of stars on subparsec to kiloparsec scales, its physical implication is not clear. In this study, we use the Illustris-TNG50 simulation to study the connection between the 2PCF of accreted halo stars and the assembly histories of Milky Way-mass galaxies. We find, in general, the 2PCF signal increases with the increase in galactocentric radii, $r$, and with the decrease in the pair separations. Galaxies which assemble late on average have stronger 2PCF signals. With $z_{1/4}$, $z_{1/2}$ and $z_{3/4}$ defined as the redshifts when galaxies accreted one-fourth, half and three-fourths of their ex-situ stellar mass at today, we find they all show the strongest correlations with the 2PCF signals at $r<\sim0.2R_{200}$. $z_{3/4}$ shows the strongest correlations than those of $z_{1/4}$ or $z_{1/2}$. However, the correlations have large scatters. The 2PCFs in velocity space show weaker correlations with the galaxy formation times within $\sim0.35R_{200}$ than real space 2PCFs, and the scatter is considerably large. Both the real and velocity space 2PCFs correlate with the assembly histories of the host dark matter halos as well. Within $0.3R_{200}$, the real space 2PCF shows stronger correlations with the galaxy formation histories than with the halo formation histories. We conclude that it is difficult to use 2PCF alone to precisely predict the formation times or assembly histories of galaxies.



\end{abstract}



\section{Introduction}
\label{sec:intro}

Two point correlation function (2PCF) and its Fourier transformed power spectrum are old but powerful statistical tools to study galaxy and matter clustering on different cosmological scales in our Universe \citep{1977ApJ...217..385G}. It has long been applied to various types of observed galaxies at different redshifts \citep[e.g.][]{2005ApJ...630....1Z,2005MNRAS.362..711Y,2005MNRAS.357..608Y,2009MNRAS.397.1862P,2011ApJ...734...88W,2012MNRAS.424.1471L,2014MNRAS.441.2398G}, and have led to significant discoveries, such as the measurement of the redshift space distortions \citep[e.g.][]{2013PhRvD..88j3510Z}, of detection of the baryon acoustic oscillations \citep[e.g.][]{2007MNRAS.381.1053P}. It has enabled halo occupations studies \citep[e.g.][]{1998ApJ...494....1J,2000MNRAS.318.1144P,2002ApJ...575..587B,2005ApJ...633..791Z,2021MNRAS.505.2784Z} and constraints on cosmological parameters \citep[e.g.][]{1998ApJ...494....1J,2005MNRAS.362..505C,2023ApJ...948...99Z}. It is a widely used technique to achieve statistical comparisons between observational data and theoretical predictions \citep[e.g.][]{2016MNRAS.456.2301W}. The cross correlations between galaxies and cosmic microwave background signals are also detected \citep[e.g.][]{2021ApJ...923..153D,2022MNRAS.511.3548S,2023A&A...673A.111Y}.

On much smaller non-linear scales, 2PCF has also been applied to quantitatively describe and study the clustering of stars. For example, on subparsec and parsec scales, it has been used to understand stellar multiplicity in star-forming regions \citep[e.g.][]{2017A&A...599A..14J,2018A&A...620A..27J,2021ApJ...922...49K}. 

From subkiloparsec to kiloparsec (kpc) scales, it has long been recognized that our Milky Way (hereafter MW) halo stars are clumpy in both real and velocity space, which are contributed by stripped stars from infalling smaller satellite galaxies or globular clusters \citep[e.g.][]{1999Natur.402...53H,2000ApJ...540..825Y,2002ApJ...569..245N,2003ApJ...588..824Y,2003MNRAS.340L..21I,2006ApJ...637L..29B,2006ApJ...642L.137B,2007ApJ...668..221N,2009ApJ...698..567S,2011ApJ...738...79X}. Stars stripped from the same progenitor satellite galaxy maintain similar orbits initially (stellar streams), which then become more phase-mixed at later stages. 2PCF can be used as a powerful statistical tool to quantify the phase-space clustering of MW halo stars and serve as the target for comparisons between observed MW halo stars and MW-like galaxies in numerical simulations. 

In an early study, \cite{2011MNRAS.417.2206C} studied mock halo star samples constructed from numerical simulations \citep{2010MNRAS.406..744C}, using 2PCFs they compared the mock with the 2PCF of observed blue horizontal branch (BHB) stars from SDSS. The simulation has no realistic disc population, but it was shown that reasonable agreement can be achieved with the signals of observed BHB stars, after placing the mock observer in a realistic position of a mock heliocentric frame and after including observational errors. In particular, their 2PCF was calculated over a 4-dimensional space formed by spatial coordinates plus the radial velocity. A metric was introduced to guarantee the same unit and similar scale between positions and the radial velocity.

With the launch of {\it Gaia} and other ground based surveys, more observational data is available. As a consequence, more and more phase-space substructures are being detected \citep[e.g.][]{2018ApJ...862..114S,2018MNRAS.478.5449M,2020ApJ...891...39Y,2019A&A...631L...9K}, including disequilibrium patterns in the Galactic disk \citep[e.g.][]{2018Natur.561..360A,2021PrPNP.12103904G}. In a more recent study, \cite{2023ApJ...942...41H} computed 2PCF signals using {\it Gaia} data release 2 data to investigate the symmetry-breaking pattern in the orthogonal and radial directions of the Galactic disk. Extensive wavelike structures are observed, indicating the system is not in steady state. Such symmetry-breaking patterns were first observed in stellar number counts of \cite{2012ApJ...750L..41W}.

It has been recognized that MW-like galaxies can have varied accretion histories, and the accretion events are imprinted in the phase-space distribution of halo stars at today \citep[e.g.][]{1993ARA&A..31..575M,2016ApJ...821....5D,2018MNRAS.474.5300D,2018NatAs...2..737D,2018Natur.563...85H,2018MNRAS.478..611B}. Since the 2PCF can quantify the phase-space clustering of halo stars statistically, it is promising to use the 2PCF to probe and see whether we can distinguish between the mass accretion or assembly histories, when comparing galaxies with different 2PCF signals at today in numerical simulations and comparing the MW to simulations.

Based on the assumption that the mass assembly histories of galaxies can be distinguished from the 2PCF of accreted halo stars at $z=0$, \cite{2019MNRAS.484.2556L} computed the 2PCF using the 3-dimensional coordinates of RR Lyrae from the Catalina Real-time Transient Survey \citep[][CRTS]{2009ApJ...696..870D} and the Panoramic Survey Telescope and Rapid Response System \citep[][Pan-STARRS1]{2016arXiv161205560C,2020ApJS..251....7F}. By comparing the measured 2PCF of observed RR Lyrae with MW-like simulations \citep{2005ApJ...635..931B}, \cite{2019MNRAS.484.2556L} picked up those simulated systems that have similar 2PCF signals as the sample of RR Lyraes. These simulated systems are thought to most closely represent the assembly histories of our MW Galaxy. 

However, it is still not clear whether the mass assembly histories can be quantitatively distinguished from the 2PCFs of halo stars. The growth of our MW stellar halo is a non-linear process, so the measured 2PCF signals often cannot be well fit by a unique functional form \citep[e.g.][]{2019MNRAS.484.2556L}. Most of the previous studies relying on numerical simulations to interpret the observed signals are empirical. 

In this study, we use the TNG50-1 hydrodynamical simulation to investigate whether we can establish quantitative connections between the 2PCF signals and the assembly histories of MW-mass galaxies. We investigate both the real and velocity space 2PCF signals, as a function of the galactocentric radii. We will show definite clues for the correlations between the 2PCF signals and galaxy assembly histories or formation times, but the scatters are very large, which prevent precise inference of the galaxy or halo formation times from the 2PCF signals of halo stars.

The layout of this paper is as follows. We introduce the TNG50-1 simulation, our selections of MW-like galaxies and halo stars in Section~\ref{sec:data}. We introduce our method of calculating the 2PCFs in Section~\ref{sec:methods}. Results will be presented in Section~\ref{sec:results}, and we summarize and conclude Section~\ref{sec:concl}. 

\section{Data}
\label{sec:data}

\begin{figure*}
\begin{center}
\includegraphics[width=0.8\textwidth]{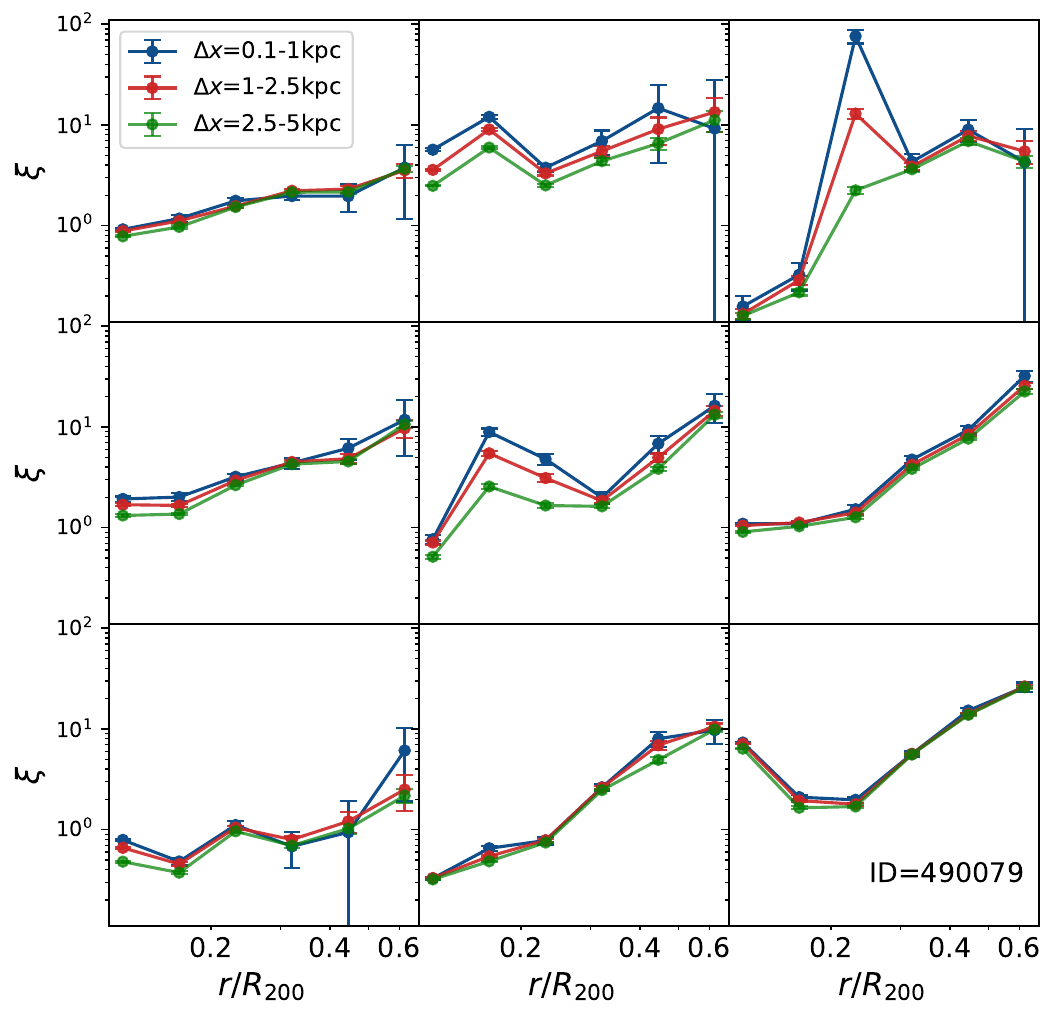}\\
\end{center}
\caption{Examples of two point correlation functions (2PCFs) for accreted halo stars of nine randomly chosen MW-mass systems from TNG50. Stars bound to satellites are not used. The $x$-axis refers to the distance to galactic center in unit of the virial radius of the host dark matter halos, $R_{200}$. Curves with difference colors refer to three different pair separations of the 2PCFs, $\Delta x=0.1-1$~kpc (dark blue), $1-2.5$~kpc (red) and $2.5-5$~kpc (green). In general, the 2PCF signals increase proportionally with the increase in the galactocentric radius, and with the decrease in $\Delta x$. The local peaks are due to substructures such as massive and dynamically cold stellar streams. Error bars are calculated from the 1-$\sigma$ scatters of 100 sub samples bootstrapped over all halo star particles in each galaxy.}
\label{fig:correxample}
\end{figure*}

\begin{figure}
\begin{center}
\includegraphics[width=0.49\textwidth]{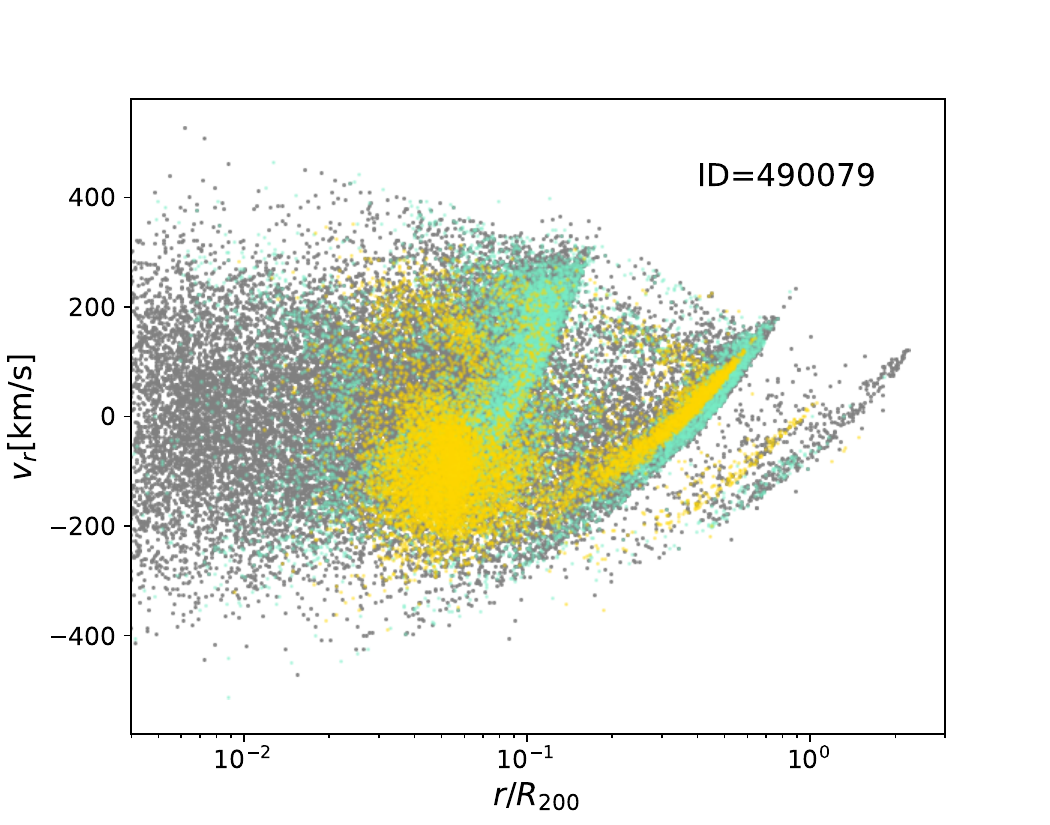}
\end{center}
\caption{Radial velocities, $v_r$, versus the galactocentric radii, $r$, for the galaxy with ID=490079, i.e., the galaxy in the bottom right panel of Figure~\ref{fig:correxample}. Colors (yellow and cyan) refer to star particles from two most massive progenitor satellites. }
\label{fig:phaseexample}
\end{figure}

Our sample of MW-mass galaxies is selected from the TNG50-1 \citep{Nelson2019,2019MNRAS.490.3234N,2019MNRAS.490.3196P}  simulation of the IllustrisTNG Project. The IllustrisTNG simulations are a suite of hydrodynamical simulations incorporating sophisticated baryonic processes, carried out with the moving-mesh code \citep[\textsc{arepo};][]{Springel2010} to solve the equations of gravity and magneto-hydrodynamics. They include comprehensive treatments of various galaxy formation and evolution processes, such as metal line cooling, star formation and evolution, chemical enrichment and gas recycling. For more details about TNG, we refer the readers to \cite{Marinacci2018,Naiman2018,Nelson2018,Pillepich2018,Springel2018,Nelson2019}.

The TNG suites of simulations adopt the Planck 2015 $\Lambda$CDM cosmological model with $\Omega_\mathrm{m}=0.3089$, $\Omega_\Lambda=0.6911$, $\Omega_\mathrm{b}=0.0486$, $\sigma_8=0.8159$, $n_s=0.9667$, and $h=0.6774$ \citep{Planck2015}. A total of 100 snapshots are saved between redshifts of $z=20$ and $z=0$, with the initial condition set up at $z=127$. TNG50-1 is the simulation with the highest resolution in its suites, and hereafter we refer to it as TNG50. It has a periodic box with 35~Mpc/h on a side that follows the joint evolution of 2160$^3$ dark matter particles and approximately 2160$^3$ baryonic resolution elements (gas cells and stellar particles). Each dark matter particle has a mass of $3.1\times10^5 \msun$/h, while the baryonic mass resolution is $5.7\times10^4 \msun$/h.

Based on the measurement of \cite{2015ApJ...806...96L}, our MW-mass systems are selected to be those central galaxies in the stellar mass range of $6.08\times10^{10\pm 0.15}\msun$ from TNG50. Moreover, in order to eliminate systems undergoing major mergers that are strongly deviating from steady state, we require the magnitudes of these central galaxies should be at least 0.5 brighter than their satellites. In the end, we select 84 MW-like galaxies\footnote{A set of MW/M31-like galaxies have been selected and described by \cite{2023arXiv230316217P}, which are publically available at https://www.tng-project.org/data/milkyway+andromeda/. The allowed stellar mass range adopted for the selection is larger than the one used in our selection to cover the mass for M31, and as we have checked, most of our MW-like galaxies are included in the TNG MW/M31-like sample. In this study, we choose to focus on our smaller sample of MW-like galaxies. }.  

For each MW-mass galaxy in our sample, we select only ex-situ formed stars\footnote{Ex-situ stars are formed by accreting smaller galaxies, while in-situ stars are formed by gas cooling.} for our analysis, and we call them halo stars or halo star particles hereafter. Moreover, we exclude star particles which are still bound to surviving subhalos or satellites. We use the stellar assembly catalogue provided by the TNG website \citep{2016MNRAS.458.2371R,Rodriguez-Gomez2017} to check whether each star particle is ex-situ or in-situ formed. This catalogue is constructed by tracking the baryonic merger trees and it is used to determine whether a stellar particle was formed outside of the ``main progenitor branch'' of a given galaxy. If true, it is considered as an ex-situ star particle. Otherwise, the star particle is tagged as an in-situ star particle. 

\section{Methodology}
\label{sec:methods}

With a continuous density field, the 2PCF, $\xi$, is defined as 

\begin{equation}
\xi(\Delta x)=\langle \delta(x)\delta(x+\Delta x) \rangle,
\label{eqn:2pcfdef1}
\end{equation}

where the average goes over the spatial coordinates, $x$, and $\delta=\frac{\rho}{\langle \rho \rangle}-1$ is the density contrast, which describes the excess in the local density, $\rho$, with respect to the mean density, $\langle \rho \rangle$. 

For a discrete density field, such as the spatial distribution of discrete galaxies or stars, the 2PCF can be equivalently written as the following form

\begin{equation}
\ud n_{12}(\Delta x)=\bar{n}_1\bar{n}_2[1+\xi(\Delta x)]\ud V_1 \ud V_2.
\label{eqn:2pcfdef2}
\end{equation}

$\xi(\Delta x)$ describes the excess probability, with respect to a uniform distribution, of finding population 2 objects (denoted by lower index 2) with a separation of $\Delta x$ to and in a volume of $\ud V_2$ around a population 1 object, denoted by lower index 1 and occupying volume $\ud V_1$. $\bar{n}_1$ and $\bar{n}_2$ are the average number densities of the two populations. $\ud n_{12}$ on the left hand side of Equation~\ref{eqn:2pcfdef2} is the pair count of population 1 and 2 objects with spatial separation of $\Delta x$. When population 1 ($\mathrm{D_1}$) and population 2 ($\mathrm{D_2}$) are identical, the 2PCFs are simply auto correlations ($\mathrm{D_1}=\mathrm{D_2}=\mathrm{D}$), and in our analysis of this paper, we calculate the auto correlations of ex-situ halo stars. 

In practice, the 2PCF is estimated from the pair counts of data points (in our case the star particles) and random points, with the random points generated following the spatial distributions of data points. In our analysis, we adopt the Landy-Szalay estimator \citep{1993ApJ...412...64L} of the following form to calculate the auto correlations

\begin{equation}
\xi(\Delta x)=\frac{\mathrm{DD}(\Delta x)-2\mathrm{DR}(\Delta x)+\mathrm{RR}(\Delta x)}{RR(\Delta x)}. 
\label{eqn:estimator}
\end{equation}

Here D stands for the sample of star particles from TNG50, and R refers to a random sample. DD, DR and RR are the total number of all possible pair counts between star particle pairs, between star particle and random point and between random pairs with a 3-dimensional separation of $\Delta x$. Throughout this paper, we use $r$ to denote the galactocentric radii, and $\Delta x$ to denote the spatial separations between the pair counts, $\Delta x=\sqrt{(x_1-x_2)^2+(y_1-y_2)^2+(z_1-z_2)^2}$. A similar separation can also be defined in velocity distance, $\Delta v=\sqrt{(v_{x,1}-v_{x,2})^2+(v_{y,1}-v_{y,2})^2+(v_{z,1}-v_{z,2})^2}$.

The random sample should trace the large scale spatial distribution of star particles, but should be free of substructures. In our case, the random sample has the same radial distribution or radial number density profile as the real halo star particles. To create the random sample, we adopt two different approaches. First, we follow the approach of \cite{2019MNRAS.484.2556L} which fit a double power law function to the radial number density profiles of halo stars in each galaxy. Based on the best-fitting double power law function, we generate random points which follow this radial distribution with random angular distributions. In the second approach, we randomly shuffle the position angles of different halo star particles. This is achieved by randomly assigning the position angle of another particle to the current particle. The two approaches lead to very similar results, and throughout this paper, we show results based on the second approach. 

When calculating the 2PCF signals in bins of different galactocentric distances, stars are binned radially first and then the 2PCF signals are calculated only among the radially binned data. The edge effect of each radial bin can be naturally accounted for, as long as the star particles and random points have exactly the same spatial coverage or bin boundary. When calculating velocity space 2PCFs in bins of different galactocentric distances, the steps are the same, but for star particles and random points in each bin, the pair counts in Equation~\ref{eqn:estimator} are calculated with pair separations defined in velocity space.

With the random sample generated following the same radial distribution as the halo stars, our measured 2PCF signals mainly represent the strength of the clumpyness in halo stars at different galactocentric radii and on different scales or pair separations. Since star particles bound to surviving subhalos or satellites are excluded, the clumpyness in our sample of halo stars reflects the existence of stellar streams. The streams approximately maintain their phase-space clusterings after getting tidally stripped from their progenitor satellite galaxies \citep[e.g.][]{2012MNRAS.420.2700K,2014ApJ...795...95B,2023MNRAS.524.2124P}. For recently stripped stellar streams, which are dynamically cold and more concentrated, their 2PCF signals will be significantly greater than unity. Our measurements of the 2PCF signals can potentially tell, on which galactocentric radii and scales are the clumpyness more dominant. Moreover, since the phase-space clustering of stellar streams becomes weaker at later times, the strength in the 2PCF signals may correlate with the time or redshift when the merger event happens. In other words, it may contain information about the mass assembly histories of the host galaxy system. 

Throughout our analysis in this paper, we adopt the open source code, \textsc{corrfunc} \citep{10.1007/978-981-13-7729-7_1,2020MNRAS.491.3022S}, to calculate the 2PCF signals. \textsc{corrfunc} is a set of fast and high-performance routines to measure clustering statistics. It supports parallel calculations and thus enables very efficient counting of the above pairs ($DD$, $DR$ and $RR$) over different pair separations or scales ($\Delta x$) in Equation~\ref{eqn:estimator}.

\section{Results}
\label{sec:results}

\subsection{2PCF in real space}

\begin{figure} 
\begin{center}
\includegraphics[width=0.49\textwidth]{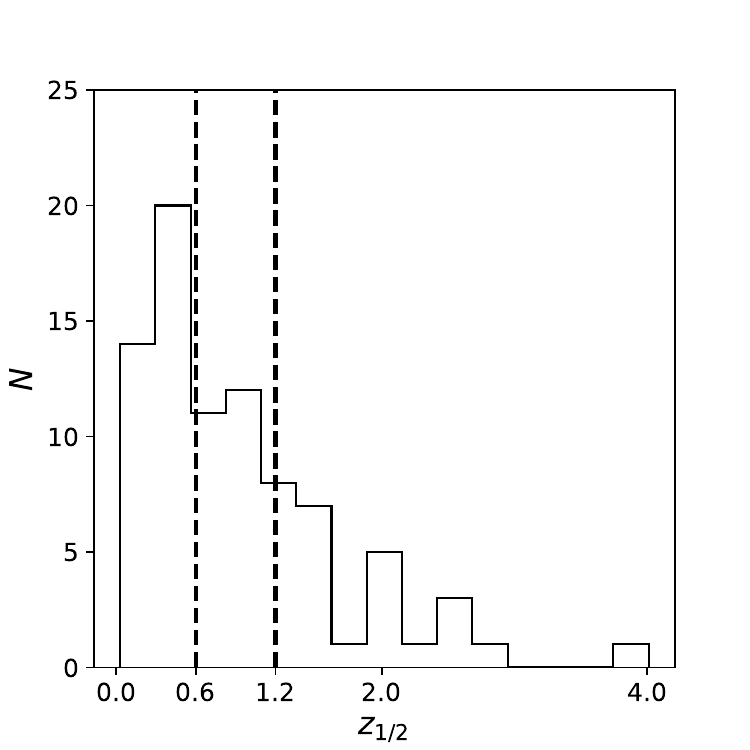}%
\end{center}
\caption{Histogram of the galaxy formation time, $z_\mathrm{1/2}$, for the ex-situ stellar mass growth of our MW-mass galaxies selected from TNG50. Here $z_{1/2}$ is defined as the redshift at which the galaxy has assembled half of its ex-situ stellar mass at $z=0$. Based on the distribution, we divide galaxies into three bins according to $z_\mathrm{1/2}<0.6$, $z_\mathrm{1/2}=0.6-1.2$ and $z_\mathrm{1/2}>1.2$ for analysis in later figures. The bin boundaries are denoted by the two black vertical dashed lines. We have similar numbers of galaxies in the three bins of $z_{1/2}$. }
\label{fig:zformhist}
\end{figure}

\begin{figure} 
\begin{center}
\includegraphics[width=0.49\textwidth]{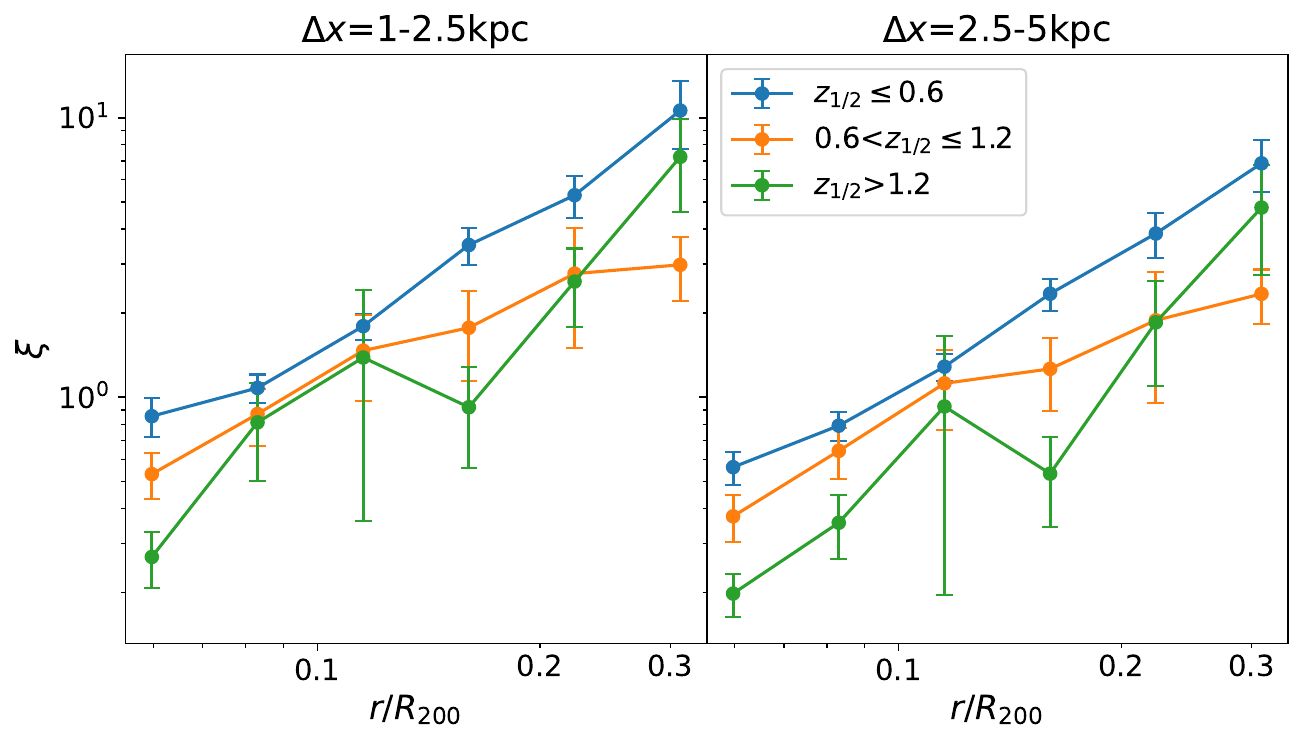}
\includegraphics[width=0.49\textwidth]{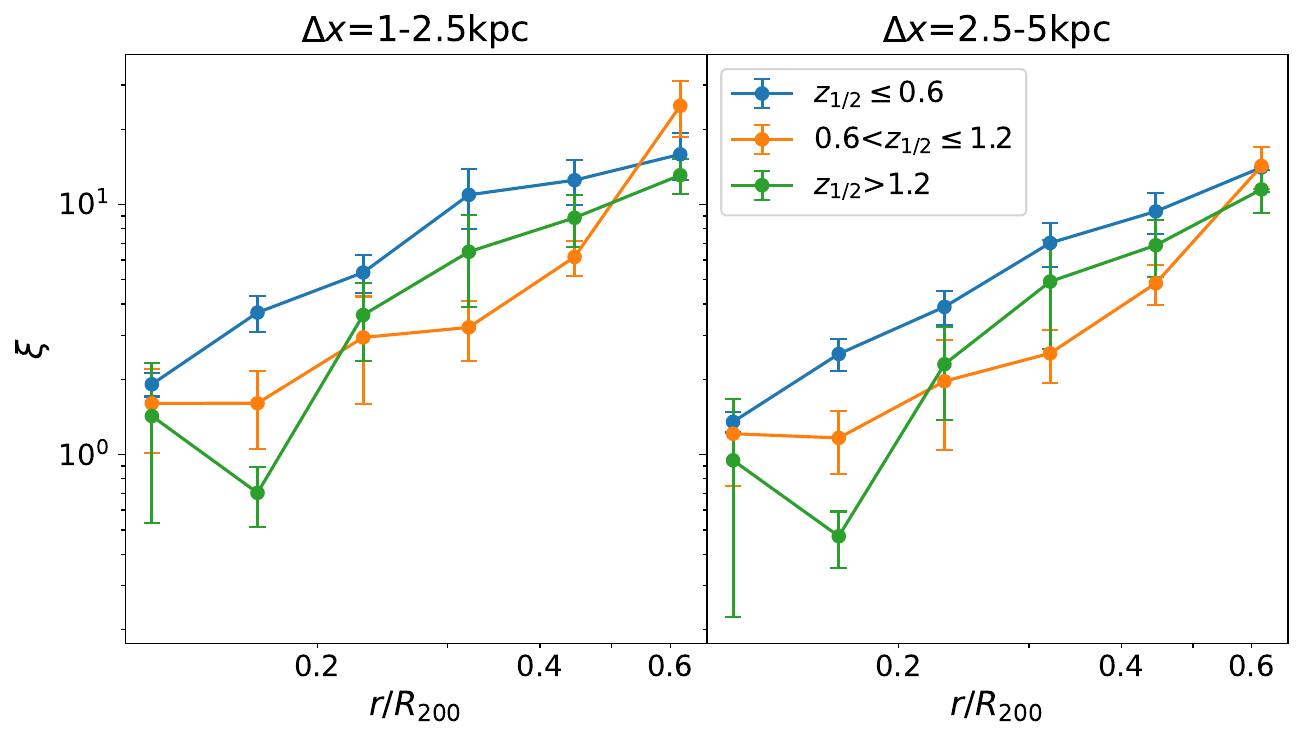}\\
\end{center}
\caption{The averaged 2PCF signals ($\xi$) of halo stars in galaxies divided into three different bins according to their galaxy formation time, $z_\mathrm{1/2}<0.6$ (blue), $z_\mathrm{1/2}=0.6-1.2$ (orange) and $z_\mathrm{1/2}>1.2$ (green). Two columns refer to pair separations of $\Delta x=1-2.5$~kpc and $\Delta x=2.5-5$~kpc used to calculate the correlation functions, as indicated by the text on top. The two top panels show results within galactocentric radii of $\sim0.3R_{200}$, with six equally spaced bins in log space. This is to highlight the results within $\sim0.3R_{200}$. The two bottom panels show the results out to $\sim0.6R_{200}$ with broader radial bins, to show that the amplitudes of the 2PCF do not change monotonically with $z_{1/2}$ beyond $r\sim0.25R_{200}$. The trend is clear in the top panels and on small scales of the bottom panels, in terms that for galaxies with later $z_\mathrm{1/2}$, the clustering amplitude in their mean two point correlation functions is stronger. Error bars are calculated from the 1-$\sigma$ scatters of 100 sub samples bootstrapped over all halo star particles in each galaxy. }
\label{fig:real3}
\end{figure}

\begin{figure*}
\begin{center}
\includegraphics[width=1\textwidth]{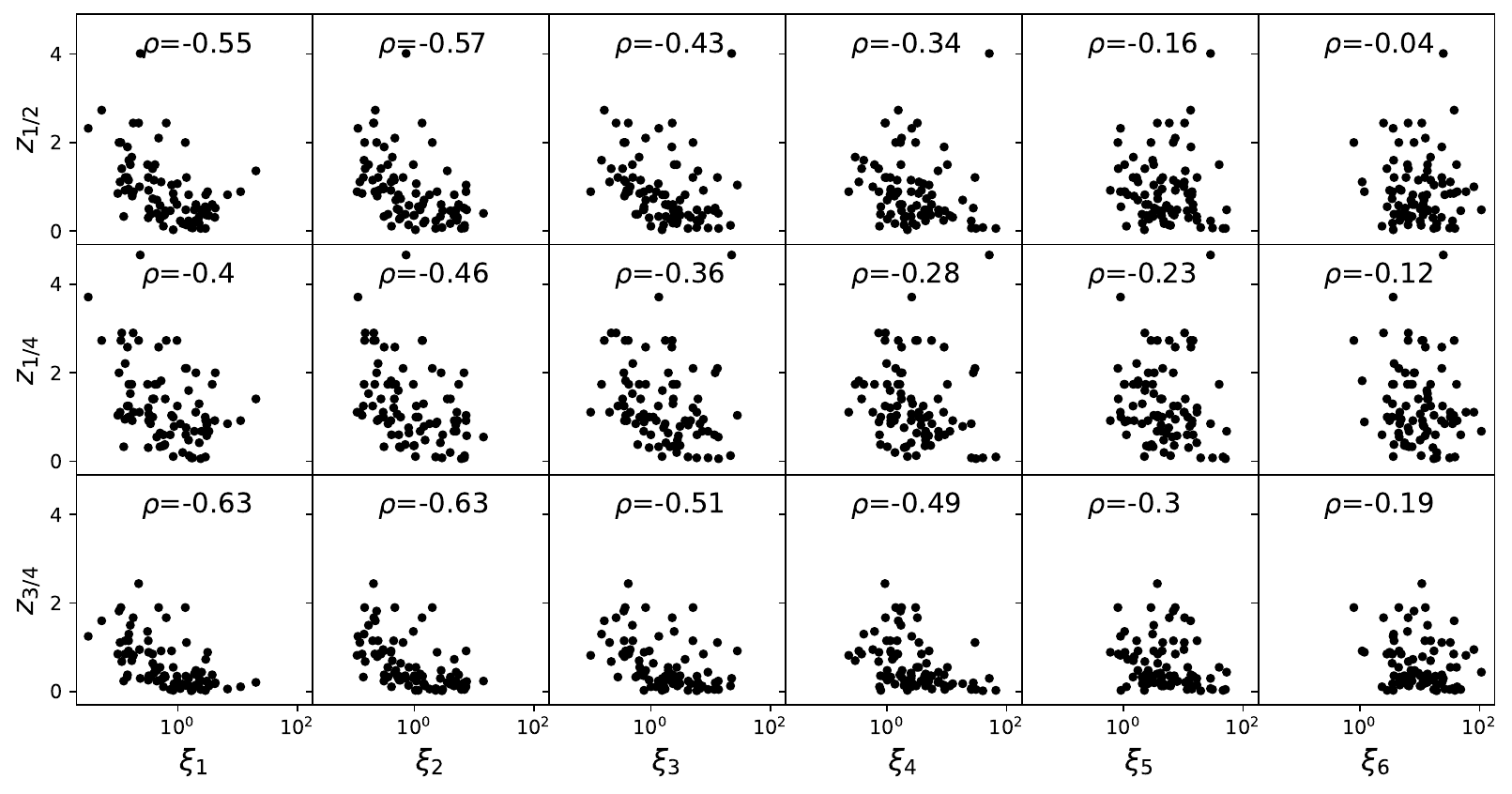}
\end{center}
\caption{Scatter plot showing the galaxy formation times defined in different ways ($z_\mathrm{1/2}$, $z_\mathrm{1/4}$ and $z_\mathrm{3/4}$) versus the 2PCF signals at different radial bins. Here $z_\mathrm{1/2}$, $z_\mathrm{1/4}$ and $z_\mathrm{3/4}$ refer to the redshifts when the galaxy has accreted half, one-fourth and three-fourths of its total ex-situ stellar mass at $z=0$. The $x$-axis refers to the 2PCF signals in the six radial bins corresponding to the $x$-axis of the bottom panels in Figure~\ref{fig:real3}, i.e., $\xi_1$ is the 2PCF signal at the smallest radial bin of the top panels in Figure~\ref{fig:real3} ($r\sim0.1R_{200}$), and $\xi_6$ is the 2PCF at the largest radial bin of the top panels in Figure~\ref{fig:real3} ($r\sim0.6R_{200}$). Each dot refers to one galaxy, and we have averaged the 2PCF signals over the measurements based on $\Delta x=1-2.5$~kpc and $\Delta x=2.5-5$~kpc. $\rho$ in each panel indicate the Spearman correlation coefficients. }
\label{fig:zvs2pcf}
\end{figure*}

\begin{figure}
\begin{center}
\includegraphics[width=0.49\textwidth]{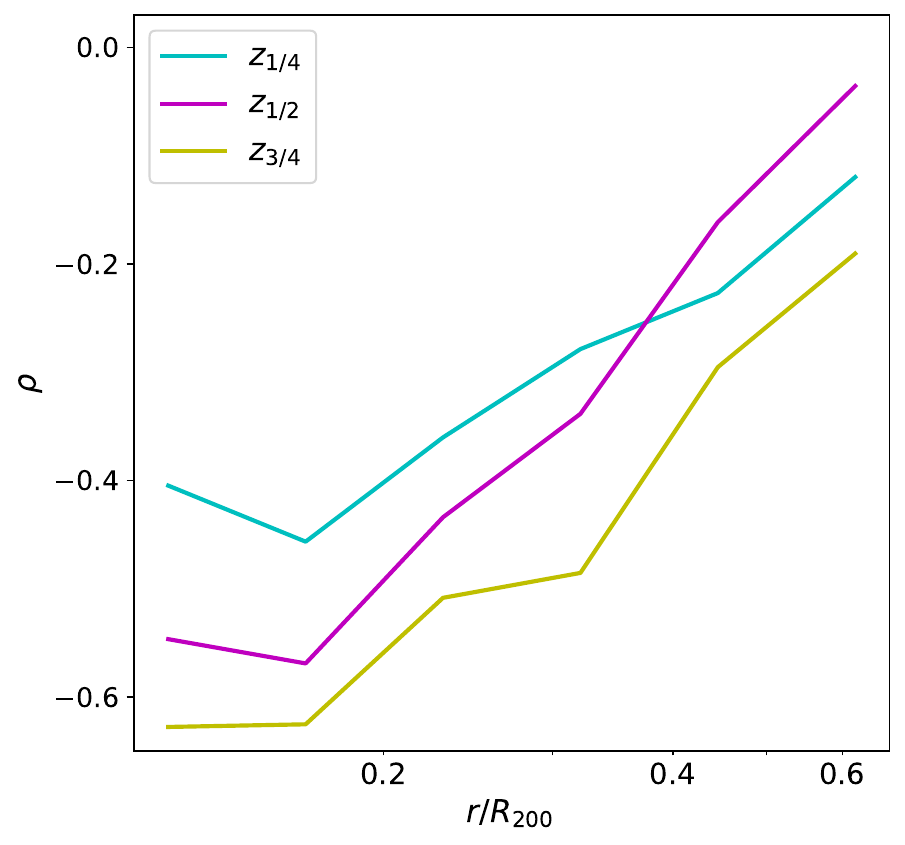}
\end{center}
\caption{Spearman correlation coefficients between the galaxy formation times and the 2PCF signals at different galactocentric radii. Cyan, magenta and yellow curves refer to the galaxy formation times, $z_\mathrm{1/4}$ , $z_\mathrm{1/2}$ and $z_\mathrm{3/4}$, respectively. }
\label{fig:coeffvsr}
\end{figure}

\begin{figure}
\begin{center}
\includegraphics[width=0.49\textwidth]{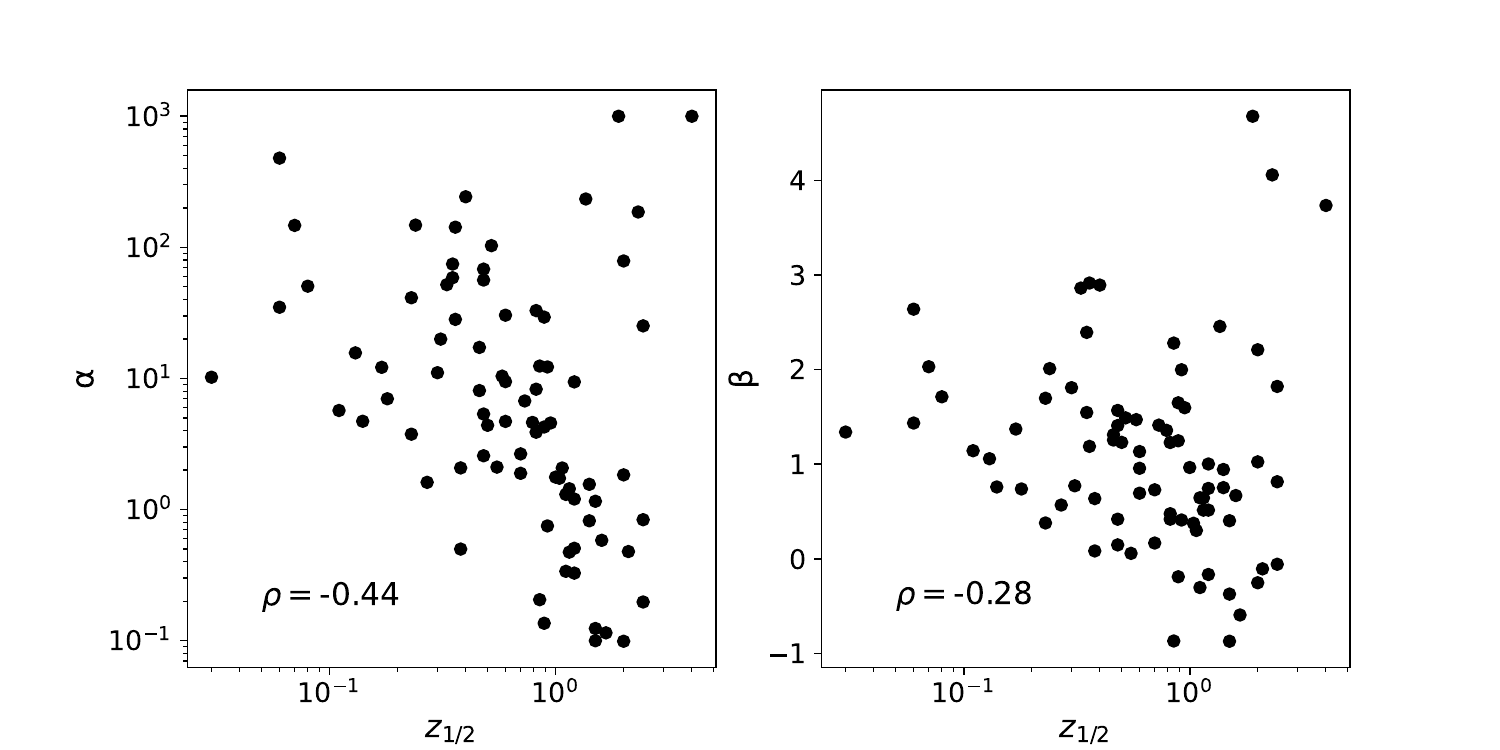}
\end{center}
\caption{We adopt a single power-law function form of $\xi =\alpha r^{\beta}$ to fit the measured 2PCFs within $0.3R_{200}$ as a function of galactocentric radii. Here $r$ is the galactocentric radius. Left and right plots show the best-fitting amplitudes and slopes versus the galaxy formation time of halo stars, $z_\mathrm{1/2}$. Each dot refers to one galaxy, and we have averaged the 2PCF signals over the measurements based on $\Delta x=1-2.5$~kpc and $\Delta x=2.5-5$~kpc. The number in each panel refers to the Spearman correlation coefficients ($\rho$).}
\label{fig:paramvsformtime}
\end{figure}

The 2PCF signals for the halo stars of nine randomly chosen MW-mass systems are shown in Figure~\ref{fig:correxample}. Unlike how the traditional 2PCFs are presented for galaxy clustering, here the $x$-axis is the galactocentric radius, $r$, normalized by the virial radius, $R_{200}$, of the host dark matter halo\footnote{The virial radius, $R_{200}$, is defined as the radius within which the mean matter density is 200 times the critical density of the universe. The total mass within $R_{200}$ is defined as the virial mass of the host dark matter halo, denoted as $M_{200}$.}. Following \cite{2019MNRAS.484.2556L}, we adopt three different pair separations of $\Delta x=0.1-1$~kpc, $1-2.5$~kpc and $2.5-5$~kpc to calculate the 2PCF, which are denoted by dark blue, red and green curves, respectively. For a given symbol at a given $r$, the signal tells the excess probability of finding pairs of halo stars over the given scales, compared with random distributions. The error bars are based on the 1-$\sigma$ scatters of 100 bootstrapped sub samples.

Overall, the clustering of halo stars, as revealed by the 2PCF signals, increases with the decrease in $\Delta x$ for most of the systems shown in Figure~\ref{fig:correxample}. Note, however, the softening scale of TNG50-1 is $\epsilon=290$~pc at $z=0$ \citep{Nelson2019,Pillepich2018}, which is greater than the lower bin boundary of the first $\Delta x$ bin ($0.1-1$~kpc). As a result, the 2PCF signals based on $\Delta x=0.1-1$~kpc (blue curves in Figure~\ref{fig:correxample}) might be affected by the resolution. Thus hereafter, we will focus our discussions on the 2PCFs averaged over the signals measured with $\Delta x=1-2.5$~kpc and $\Delta x=2.5-5$~kpc, but given the small difference between the different color curves in Figure~\ref{fig:correxample}, our conclusions are not affected even if we include the signals based on $\Delta x=0.1-1$~kpc or if we present results based on the three different $\Delta x$ separately. Moreover, we note the choice of $\Delta x$ does not have conformity when calculating the 2PCFs for stars. In a previous study, \cite{2023ApJ...942...41H} have performed detailed investigations on different bin width and the limiting scale length that can be probed, showing how the 2PCF signals change with the pair separations. In our analysis, the 2PCFs are computed with relatively wider pair separations than those explored in \cite{2023ApJ...942...41H}. However, our 2PCFs do not show strong changes with different choices of $\Delta x$, and in most cases the 2PCF signals increase monotonically with the decrease in $\Delta x$ with reasonable errorbars, and thus our conclusions barely change if we vary the bin width of $\Delta x$ within a reasonable range.

The 2PCF signals prominently increase with the increase of galactocentric radii. This is because at larger radii, halo stars were stripped from their progenitor satellites more recently, which preserve better the initial clustering and hence have stronger 2PCF signals. At later times, satellite galaxies and stellar streams fall to more central regions of the host dark matter halos due to dynamical frictions, and they also gradually become less clustered in phase space. Thus the 2PCF signals are weaker at smaller galactocentric radii. 

A few systems in Figure~\ref{fig:correxample} show non-monotonic changes in their 2PCF signals as a function of $r$, with prominent dips and bumps, such as the upper middle, middle middle and the lower right panels. This is due to the existence of some local massive and dynamically cold streams, which contribute to the local bumps or increases in the 2PCF signals. For example, we show in Figure~\ref{fig:phaseexample} the scatter plot of radial velocity, $v_r$, versus the galactocentric distance, $r$, for halo stars in the galaxy referring to the bottom right panel of Figure~\ref{fig:correxample}. Particles from a few most massive progenitor satellites are marked in color. Particles with the same color come from the same progenitor. By comparing the bottom right panel of Figure~\ref{fig:correxample} and Figure~\ref{fig:phaseexample}, we can see when the 2PCF signals increase towards $r\sim0.1R_{200}$ and $r\sim0.6R_{200}$, there are also very prominent and coherent streams.

Note when selecting our sample of MW-mass galaxies, we have required that the central galaxies should be at least 0.5 mag brighter than their satellites. This has eliminated some systems which are undergoing major mergers, hence removed systems with dips and bumps in their 2PCF signals. However, a few systems satisfy the selection criteria at $z=0$, but have just undergone prominent merger events. The merged massive satellite has been mostly or completely disrupted, so the system passed our selection, but the resulting massive and dynamically cold streams cause such prominent dips and bumps in their 2PCF signals. The galaxy formation time of such systems, as we have checked, are on average more recent. Since we need enough number of MW-mass systems to cover a wide range of different mass assembly histories, we do not explicitly remove such systems from our sample. 

\subsection{Connection between the real space 2PCF and the stellar mass assembly history}

We now investigate how are the 2PCF signals related to the stellar mass assembly histories of galaxies. We first perform an estimate of the general trend. In Figure~\ref{fig:zformhist}, we show the distribution of the galaxy formation time, $z_\mathrm{1/2}$, for our sample of MW-mass galaxies. Here $z_\mathrm{1/2}$ is defined as the redshift when the galaxy has accreted half of its total ex-situ stellar mass at $z=0$. Note our mass assembly histories and formation times in this paper are all based on only ex-situ stars, not including in-situ stars. We make this choice because the phase-space clustering of accreted halo stars is expected to correlate more with the formation times of also the accreted part of stellar mass. In addition, we have checked that including in-situ stars would weaken the correlation. According to the distribution of $z_\mathrm{1/2}$ shown in Figure~\ref{fig:zformhist}, we divide our sample of galaxies into three bins, $z_\mathrm{1/2}<0.6$, $z_\mathrm{1/2}=0.6-1.2$ and $z_\mathrm{1/2}>1.2$. 

For galaxies even in the same bin of $z_\mathrm{1/2}$, their 2PCFs show very large scatters, and thus instead of showing the results for individual galaxies, we show in Figure~\ref{fig:real3} the mean 2PCF signals averaged for galaxies in the three bins of $z_\mathrm{1/2}$ (different color curves, see the legend). For the bottom panels, we show the 2PCFs out to $\sim0.6R_{200}$, while the upper panels highlight the signals within $\sim0.3R_{200}$. In other words, the only differences between the top and bottom panels are their $x$-axes ranges and the binning in galactocentric radii. The two panels in the same row refer to two different pair separations, $\Delta x$, as indicated by the text on top. Within $\sim0.25R_{200}$, we can see the averaged signals show significant monotonic trend, in terms that galaxies with later $z_\mathrm{1/2}$ tend to have stronger 2PCF signals on average. This is because for galaxies which assemble late, the accreted stars have less time to be phase mixed, hence resulting in stronger phase-space clustering and 2PCF signals. However, beyond $\sim0.25R_{200}$, the trend is no longer monotonic. The orange points have the lowest amplitude over $\sim0.25-0.5R_{200}$, whereas it has the highest amplitude at $\sim0.6R_{200}$, despite the fact that the corresponding $z_{1/2}$ bin for the orange curve is in between of the blue and green curves.

We show in Figure~\ref{fig:zvs2pcf} the 2PCF signals measured at different galactocentric radius bins, versus the galaxy formation times. Here we choose to show the galaxy formation times defined in a few different ways. Along the $y$-axis, $z_{1/2}$, $z_{1/4}$ and $z_{3/4}$ refer to the redshifts when the galaxy has accreted half, one-fourth and three-fourths of its total ex-situ stellar mass at $z=0$. $\xi_{1-6}$ refer to the 2PCF signals measured at the smallest to the largest galactocentric radii in the bottom panels of Figure~\ref{fig:real3}, i.e., from $\sim$0.1$R_{200}$ to $\sim$0.6$R_{200}$. Each data point refers to the 2PCF signal for one galaxy, and for the signal in a given galactocentric radius bin of each galaxy, we have averaged the 2PCF signals for the measurements based on $\Delta x=1-2.5$~kpc and $\Delta x=2.5-5$~kpc.  The numbers in each panel indicate the Spearman correlation coefficients. We can see $z_{1/4}$, $z_{1/2}$ and $z_{3/4}$ all show the strongest correlations with the 2PCF in the first and second columns, i.e., $r<\sim0.2R_{200}$. This is consistent with Figure~\ref{fig:real3}. In the bottom panels of Figure~\ref{fig:real3} we can see for galaxies in bins of later formation times, they on average have stronger 2PCF signals only within $\sim0.25R_{200}$. Now we show that the Spearman correlation coefficients are also the strongest (most negative) within $\sim0.2R_{200}$. At larger radii, the correlation becomes weaker, and we fail to see monotonic trends between the 2PCF signal strength and the formation times in Figure~\ref{fig:real3} as well, i.e., the green curves with the earliest formation times have higher amplitudes than the orange curve with the intermediate formation times in a few outer data points.

For a direct comparison, we show in Figure~\ref{fig:coeffvsr} the Spearman correlation coefficients as a function of galactocentric radii, and out to $r\sim0.6R_{200}$. Cyan, magenta and yellow curves refer to the correlation coefficients between the 2PCF signals and $z_{1/4}$, $z_{1/2}$  and $z_{3/4}$, respectively. $z_{3/4}$ quantifies the fraction of mass accreted at later redshifts than those quantified by $z_{1/2}$ or $z_{1/4}$. As a consequence, $z_{3/4}$ shows the strongest correlations (the most negative) with the 2PCF signals than those of $z_{1/2}$ or $z_{1/4}$ at every radii. This is expected, because for halo stars that are accreted later, they on average preserve better phase-space clusterings, causing stronger correlations between the 2PCF signals and the galaxy formation histories. The correlations are also stronger for $z_{1/2}$ than that of $z_{1/4}$ within $0.36R_{200}$, but beyond $0.36R_{200}$ the trend switches. This is perhaps because the majority of mass as quantified by $z_{1/4}$ is currently in the central regions, resulting in weaker correlations between $z_{1/4}$ and the 2PCF signals at large galactocentric radii. 

Figures~\ref{fig:real3}, \ref{fig:zvs2pcf} and \ref{fig:coeffvsr} all show that the correlation between the galaxy formation times and the 2PCF signals is the strongest at $r<\sim0.2R_{200}$. This is because the number of halo stars is larger in more central regions. The density profiles of accreted halo stars drop very quickly with the increase in galactocentric radii, so particles on smaller scales dominate the determination of the global galaxy formation times.

We can see in all panels of Figure~\ref{fig:zvs2pcf}, however, the scatters are large. We then adopt a single power law functional form 

\begin{equation}
    \xi=\alpha r^\beta,
\end{equation}

to fit the 2PCF signals within $0.3R_{200}$ as a function of the galactocentric radii and for each individual galaxy. In Figure~\ref{fig:paramvsformtime}, we show the best-fitting amplitude ($\alpha$) and power-law index ($\beta$) versus $z_{1/2}$. Each dot refers to one galaxy, and we have averaged the results over the measurements with $\Delta x=1-2.5$~kpc and $\Delta x=2.5-5$~kpc. It is clear that with later $z_{1/2}$, $\alpha$ and $\beta$ get larger, but there are large scatters. The scatter is mainly due to intrinsic reasons which we discuss below, but might be partly related to the reason that some of the 2PCF signals as a function of the galactocentric radii cannot be perfectly fit by single power laws. 

Figures~\ref{fig:real3}, \ref{fig:zvs2pcf}, \ref{fig:coeffvsr} and \ref{fig:paramvsformtime} show us that there are correlations between the mass assembly history of galaxies and the 2PCF signals of accreted halo stars. However, the amount of scatters is not negligible. The scatter reflects the limiting precision of using 2PCFs to constrain the mass assembly histories of galaxies. In other words, the phase-space clustering of halo stars is also affected by many other aspects, such as the infalling orbits of the progenitors. Besides, halo stars which are fully phase mixed might have lost the information about the accretion event. In order to fully understand this, we move on to carry more detailed investigations between the 2PCF signals and the mass assembly histories. We calculate the $\chi^2$ between the 2PCF signals of every possible combinations of two galaxies, and check for galaxies with very similar or very different 2PCF signals of their halo stars, how similar or how different are their mass assembly histories. 

\begin{figure*}
\begin{center}
\includegraphics[width=1\textwidth]{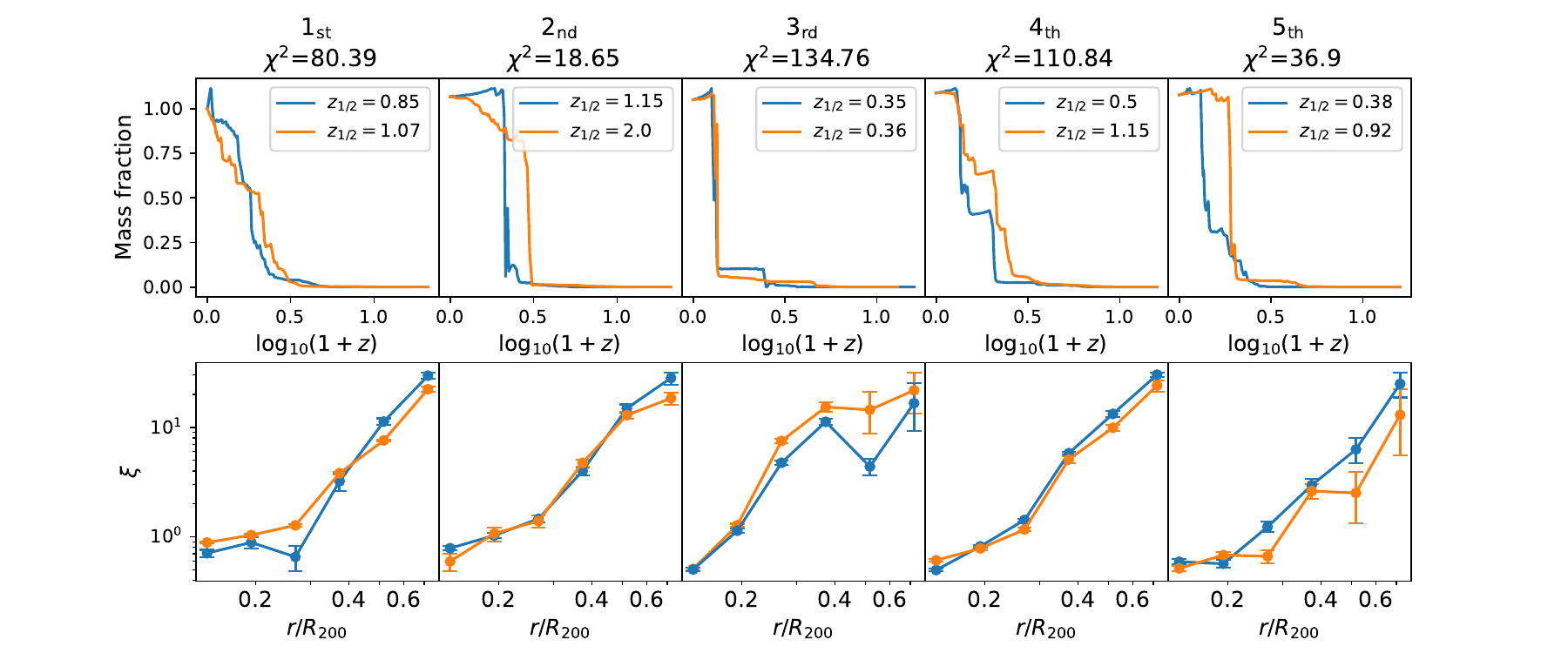}
\end{center}
\caption{Five galaxy pairs which have similar 2PCF signals (bottom row), and the corresponding mass assembly histories are shown in the top row, with the galaxy formation time, $z_{1/2}$, shown by the legend. Note when plotting the mass assembly histories, only ex-situ stellar masses are used.  Blue curves refer to the galaxy with later $z_{1/2}$, while orange curves refer to the galaxy with earlier $z_{1/2}$. The $\chi^2$ value between the 2PCFs of each galaxy pair is shown on top of each column. Error bars of the 2PCFs are calculated from the 1-$\sigma$ scatters of 100 sub samples bootstrapped over all halo star particles in each galaxy. }
\label{fig:2pcfsimilar}
\end{figure*}

\begin{figure*}
\begin{center}
\includegraphics[width=1\textwidth]{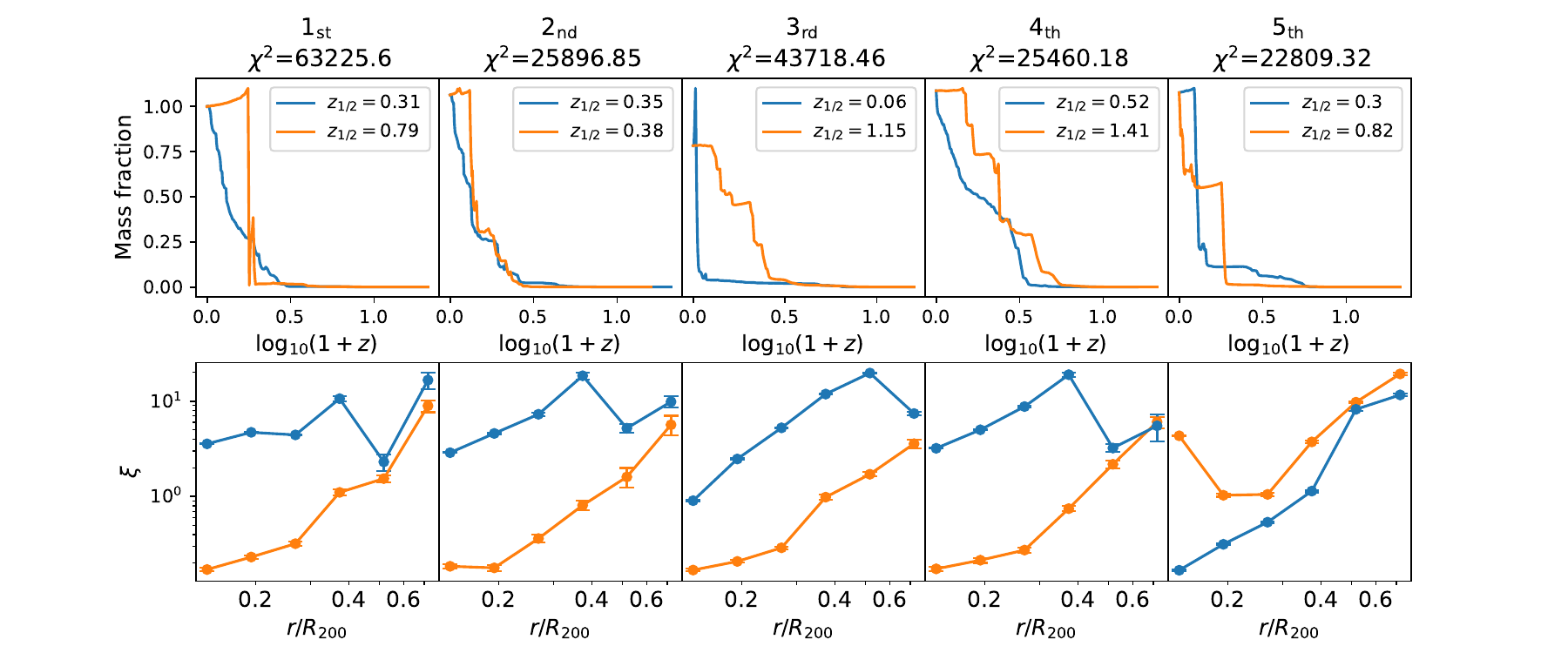}
\end{center}
\caption{Similar to Figure~\ref{fig:2pcfsimilar}, but for five galaxy pairs which have more different 2PCF signals. The $\chi^2$ value between the 2PCFs of each galaxy pair is shown on top of each column. }
\label{fig:2pcfdiff}
\end{figure*}

\begin{figure}
\begin{center}
\includegraphics[width=0.49\textwidth]{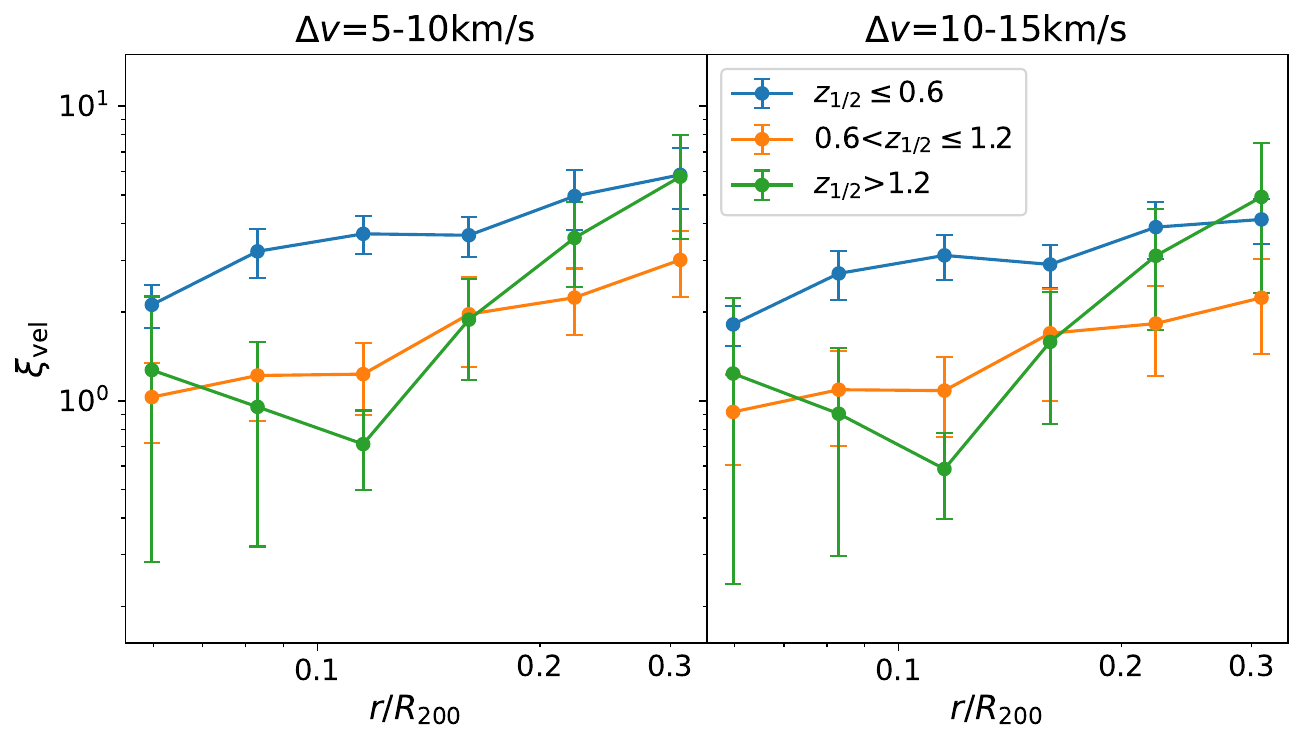}
\end{center}
\caption{Similar to the top panels of Figure~\ref{fig:real3}, but shows the two point correlation function signals in velocity space. Different color curves are averaged 2PCFs of MW-mass galaxies in three bins of formation time ($z_{1/2}$, see the legend). To distinguish from the real space 2PCFs, we denote the $y$-axis with $\xi_\mathrm{vel}$. The two panels refer to two different pair separations of halo stars in velocity space ($\Delta v$), as indicated by the text on top. Error bars are calculated from the 1-$\sigma$ scatters of 100 sub samples bootstrapped over all halo star particles in each galaxy. We choose not to show the results out to larger galactocentric radii, as the correlations between the 2PCF signals and the formation times are very weak there.}
\label{fig:vel3}
\end{figure}

\begin{figure*} 
\begin{center}
\includegraphics[width=1\textwidth]{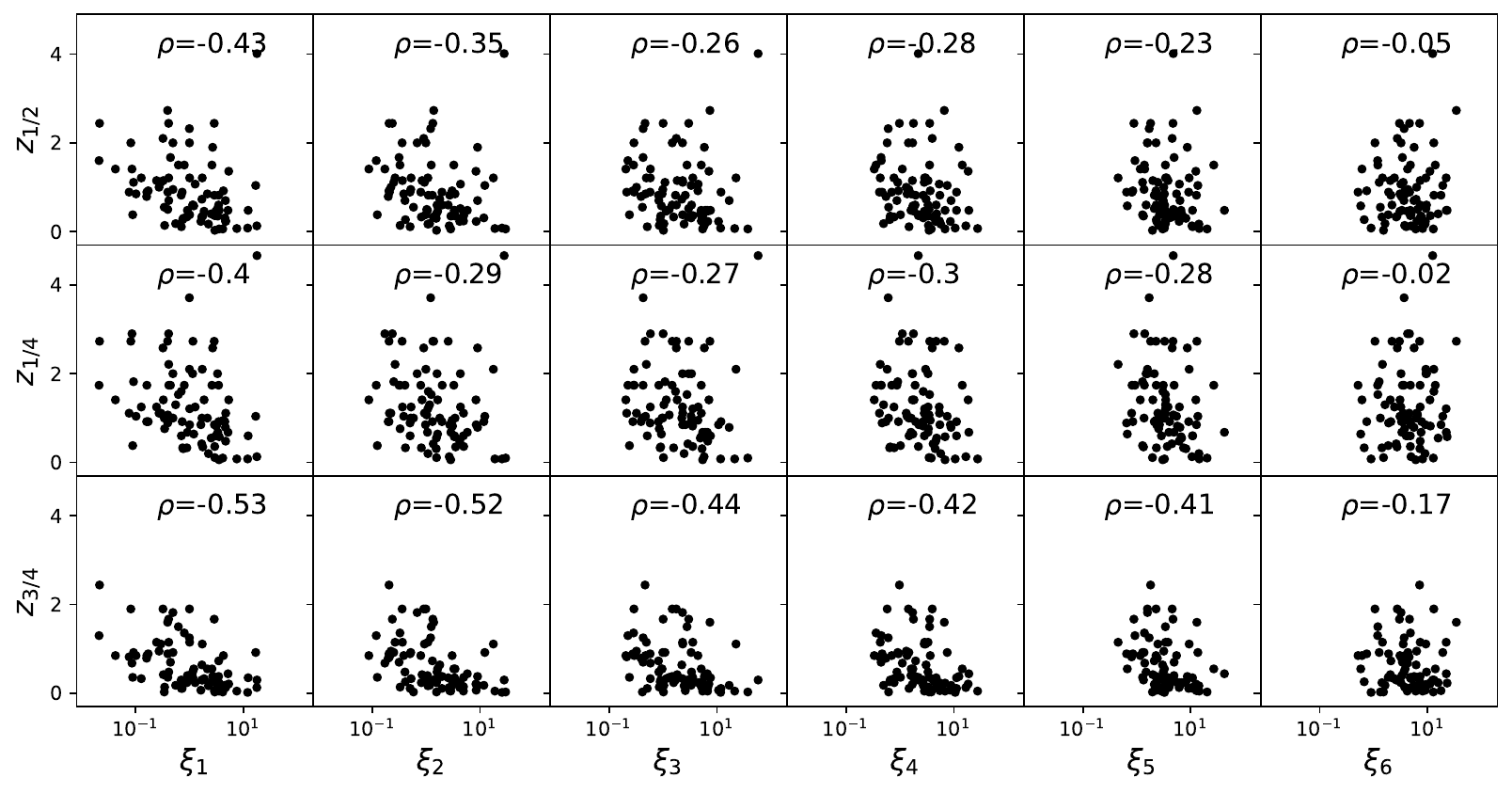}
\end{center}
\caption{Similar to Figure~\ref{fig:zvs2pcf}, but shows the results for 2PCF signals in velocity space. $\xi_1$ to $\xi_6$ refer to the 2PCF signals in six equal log spaced bins from $\sim0.1R_{200}$ to $0.6R_{200}$.
}
\label{fig:zvs2pcfvel}
\end{figure*}

\begin{figure*}
\begin{center}
\includegraphics[width=1\textwidth]{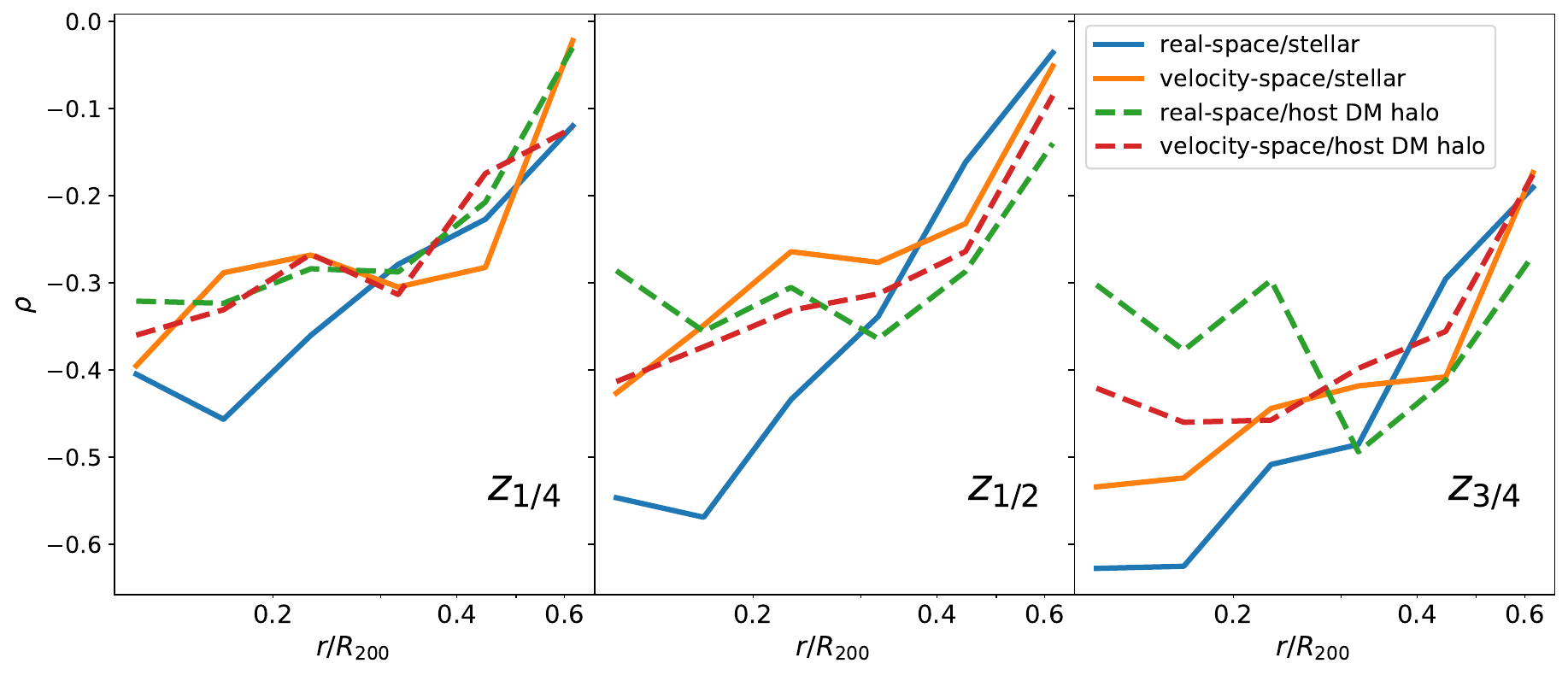}
\end{center}
\caption{Spearman correlation coefficients between the galaxy formation times and the 2PCF signals at different galactocentric radii ($y$-axis), reported as a function of the galactocentric radii ($x$-axis). Panels from the left to the right refer to formation times defined as the redshifts when the galaxies or host dark matter halos have accreted one-fourth, half and three-fourths of their total accreted mass at today ($z_\mathrm{1/4}$, $z_\mathrm{1/2}$ and $z_\mathrm{3/4}$). In all panels, blue/green and orange/red curves are based on the 2PCF signals calculated in real and velocity space, respectively. Blue and orange curves are based on the formation times of the galaxies, while green and red dashed curves are based on the formation times of the host dark matter halos. }
\label{fig:zm200vs2pcf}
\end{figure*}

We calculate the $\chi^2$ values of the 2PCF signals between every two different galaxies in our sample based on the following equation:

\begin{equation}
\chi^2=\sum_i \frac{(\xi_{i,\mathrm{galaxy1}}-\xi_{i,\mathrm{galaxy2}})^2}{\sigma_{i,\mathrm{galaxy1}}^2+\sigma_{i,\mathrm{galaxy2}}^2},
\label{eqn:chi2}
\end{equation}

where the summation goes over the 2PCF signals at different galactocentric radius bins, denoted by $i$. In the denominator, $\sigma_\mathrm{galaxy1}$ and $\sigma_\mathrm{galaxy2}$ are the errors in the 2PCFs of the two galaxies, which are computed from the 1-$\sigma$ scatters of 100 bootstrapped sub samples.

Figure~\ref{fig:2pcfsimilar} shows five examples, in which the 2PCF signals of the two galaxies under consideration in each panel are similar to each other (bottom row). Given the similarity in their 2PCF signals, the actual mass assembly histories are similar. Note for the mass assembly histories in the top row, only ex-situ stellar masses are used. The sharp changes in the mass assembly histories would not exist, if we plot them using the total stellar mass, but the correlations between the trends in mass assembly histories and the 2PCFs become weaker after including the in-situ formed stellar mass.

We can see some reasonable correlations between the top and bottom panels of Figure~\ref{fig:2pcfsimilar}. For example, in the first and second columns, for the galaxies which have later $z_{1/2}$ (blue curve), the corresponding 2PCF signals are higher in amplitude than the other one (orange) beyond $\sim0.4R_{200}$. However, the trend switches or becomes less clear at smaller radii, in terms that the galaxies with later $z_{1/2}$ (blue) have lower or similar 2PCF signals to the other one on such smaller radii. This is perhaps because the other galaxy (orange) has more matter assembled at earlier redshifts\footnote{The orange curves are above the blue ones at $\log_{10}(1+z)>\sim0.3$ in the two upper left panels.}, which sink to smaller galactocentric radii due to dynamical friction, piling up more over radial scales of 0.1 to 0.35~$R_{200}$ and hence increasing the 2PCF signals there. Note in previous figures we have shown that $z_{1/2}$ {\it on average} has stronger correlations with the 2PCF signals at smaller radii, but now the two individual cases in the first and second columns of Figure~\ref{fig:2pcfsimilar} show that the 2PCF signals at larger radii beyond $\sim0.4R_{200}$ is more consistent with the difference in $z_{1/2}$. This again reflects large system to system scatters in our analysis.

In the fourth and fifth columns, for the galaxy which assemble later (the blue curve) and hence has a much later formation time than the other one, we can see the corresponding 2PCF signals are higher in amplitudes, with the blue curves having higher amplitudes than the orange ones. However, in the fourth example, we may expect a larger difference in the 2PCF signals given the more prominent difference in their assembly histories or formation times in the upper panel. 

The third column of Figure~\ref{fig:2pcfsimilar} is a relatively bad example. The mass assembly histories of the two galaxies under consideration are quite similar to each other. The blue curve has a slightly later formation time, but in the lower panel the blue curve is lower in amplitude. It means that the galaxy which formed slightly later has lower clustering amplitude. However, the differences in either the mass assembly histories or the 2PCFs are not very large, and given the large scatters in previous figures, it is not surprising to see such a case. 

In Figure~\ref{fig:2pcfdiff}, we show five example galaxy pairs which have significantly larger differences in their 2PCF signals. In the first, second and fifth columns, when their 2PCF signals are significantly more different than those in Figure~\ref{fig:2pcfsimilar}, their mass assembly histories are not as significantly different. In particular, the fifth example is an extreme case. The orange galaxy has a much stronger 2PCF signal than the other galaxy, but it forms earlier. 

The examples presented in Figures~\ref{fig:2pcfsimilar} and \ref{fig:2pcfdiff} demonstrate the large scatter when using the 2PCF signals to infer the mass assembly history or formation time, which is likely determined by several factors such as the infalling orbits of the progenitor satellites. Nevertheless, the general trends are reasonable in most cases. In most cases, the galaxy which assembles earlier also has lower 2PCF signals, except for the third column of Figure~\ref{fig:2pcfsimilar} and the fifth column of Figure~\ref{fig:2pcfdiff}. For the fifth case of Figure~\ref{fig:2pcfdiff}, in fact halo stars in the galaxy with later formation time has significantly more radial orbits than those in the other galaxy, causing much faster phase mixing and thus lower amplitude in its 2PCF signal.

\subsection{Connection between the velocity space 2PCF and the stellar mass assembly history}

Our results in the previous subsections show that the real space 2PCF signals correlate with the stellar mass assembly histories of MW-mass galaxies, but the scatters are considerably large. In this subsection, we investigate correlations in velocity space. We check whether the velocity space 2PCF can provide more information.

Figure~\ref{fig:vel3} is similar to the top panels of Figure~\ref{fig:real3}. The $x$-axis is the galactocentric radius, $r$, scaled by the virial radius, $R_{200}$. The blue, orange and green curves are the average 2PCF signals for galaxies in three bins of $z_{1/2}$. We calculate the 2PCF in velocity space. The two columns refer to two different pair separations in velocity space, as indicated by the text on the top of each panel. Note the choice of the pair separations in velocity space ($\Delta v$) is less well defined in the literature, and here we have carefully tried many different choices of $\Delta v$, and the chosen values of $\Delta v$ are determined by the cases when we can see the most prominent monotonic trend that the 2PCF signal strength increases with the decrease in $z_{1/2}$. 

The 2PCF signal amplitudes only monotonically increase with the decrease in $z_{1/2}$ for the second and third data points counting from the center, but not at other radii. Compared with Figure~\ref{fig:real3}, it seems that the correlations in the velocity space 2PCF and the galaxy formation times are weaker than those revealed in the real space 2PCF signals. We do not show the 2PCF signals calculated out to larger galactocentric radii as in the bottom panels of Figure~\ref{fig:real3}, because on such large scales the trends are not monotonic.

We also show the correlations between the velocity space 2PCF signals at different galactocentric radii and the galaxy formation times in Figure~\ref{fig:zvs2pcfvel}. The Spearman correlation coefficients are indicated in each panel of Figure~\ref{fig:zvs2pcfvel}. Compared with the real space 2PCF (Figure~\ref{fig:zvs2pcf}), the Spearman correlation coefficients get closer to zero for $\xi_\mathrm{vel,1-3}$. In order to have direct comparisons, we plot the Spearman correlation coefficients based on real space and velocity space 2PCF signals together in Figure~\ref{fig:zm200vs2pcf}, as a function of the galactocentric radii. Comparing the blue and orange solid curves, we can see the real space 2PCF signals show stronger (more negative) correlations with the galaxy formation times at $r<\sim0.3-0.35R_{200}$ than the velocity 2PCF signals.  This is consistent with the fact that we see weaker correlations in Figure~\ref{fig:vel3} than Figure~\ref{fig:real3}. At $r>0.35R_{200}$, the velocity space 2PCF signals show slightly stronger correlations. We thus conclude that the velocity space clustering is slightly better preserved at large galactocentric radius ($r>\sim0.35R_{200}$) than the real space correlation functions. However, the scatters in Figure~\ref{fig:zvs2pcf} are very large, and thus using velocity space 2PCF does not help to significantly improve the accuracy of using real space 2PCF signals to predict the galaxy assembly histories.

\subsection{Connection between the 2PCF and the halo mass assembly history}

So far, we have investigated the correlation between the real and velocity space 2PCFs and the assembly history of the ex-situ stellar mass for MW-mass galaxies. We can also study the connection between the 2PCFs of halo stars and the assembly history of the host dark matter halo. Since both the host dark matter halos and the halo stars form through accretions of smaller halos and galaxies, and dark matter leaves imprints on the visible matter distribution \citep[e.g.][]{2021PrPNP.12103904G,2021PhRvL.126i1101N,2021ApJ...920L..11N}, we expect the phase-space clustering and the 2PCF signals of halo stars carry information about the assembly history of the host dark matter halos. In this subsection, we examine how the 2PCF signals of halo stars are related to the mass assembly history of the host dark matter halo.

In Figure~\ref{fig:zm200vs2pcf}, we show the correlation coefficients based on the real space and velocity space 2PCFs, and based on the formation times of ex-situ halo stars in galaxies and the host dark matter halos together. The green and red dashed curves use the formation times of the host dark matter halos, to be compared with the blue and orange solid curves. The green and red dashed curves show negative correlations between the halo star 2PCF signals and the halo formation times. Moreover, the velocity space 2PCFs do not show significant differences in their correlations with the galaxy or halo formation times, given the fact that the orange solid and red dashed curves show similar trends and amplitudes. This tells us that the velocity space 2PCF signals have similar performance in predicting the galaxy or halo formation histories. However, the blue solid and green dashed curves (real space 2PCFs) are more different. At $r<0.3R_{200}$, the blue solid curves are significantly below the green dashed curves, implying stronger negative correlations between the real space 2PCF signals and the galaxy formation times than that of the halo formation times. At $r>0.3R_{200}$, the blue solid and green dashed curves become more similar, with the green dashed curves slightly below the blue solid ones in the middle and right panels. Our results show that the clustering or 2PCF signals of halo stars are correlated with both the galaxy and halo assembly histories. In inner regions ($r<0.3R_{200}$) the real space 2PCF has a stronger correlation with the galaxy formation time than the halo formation time. 

\section{Discussions and conclusions}
\label{sec:concl}

The two point correlation function (2PCF) has been used to measure the spatial clustering of galaxies and dark matter halos, in real observation and in numerical simulations. In addition, the cross correlations between different types of galaxies, between galaxies and dark matter, and between galaxies and other multi-wavelength observations are powerful statistical tools to help understanding galaxy formation physics and cosmology. 

While 2PCF has been particularly useful to study galaxies and cosmology, there are also attempts of using 2PCF statistics to study the clustering of stars on subparsec to kiloparsec scales \citep[e.g.][]{2017A&A...599A..14J,2018A&A...620A..27J,2021ApJ...922...49K,2011MNRAS.417.2206C,2019MNRAS.484.2556L}. However, unlike the clustering of galaxies, the physical mechanisms that drive the clustering and dissolving of stars are non-linear, and thus, the interpretation of the 2PCF signals of stars is often empirical.

In this study, we investigate the connection between the 2PCF signals of accreted halo stars in Milky Way-mass galaxies on kpc scales and the assembly histories of these galaxies. We use the cosmological hydrodynamical TNG50 simulation. Halo stars belonging to any surviving subhalos/satellites are not used, and our 2PCF mainly reflects the strength of the clumpyness in halo stars due to the existence of stellar streams. We investigate the 2PCF signals in both real space and velocity space, and at different galactocentric radii. In general, the 2PCF amplitude increases with the decrease in pair separations, and increases proportionally with the increase in galactocentric radii, because the clustering of phase-space structures preserve better on smaller pair separations and on larger galactocentric radii. In addition, the existence of dynamically cold and massive stellar streams can cause local increases in the 2PCF amplitudes. 

We find evidences that there are correlations between the 2PCF signals and the galaxy formation times. We define the redshifts at which the galaxy accreted one-fourth, half and three-fourths of its total ex-situ stellar mass at $z=0$, as $z_{1/4}$, $z_{1/2}$ and $z_{3/4}$. All formation times show negative correlations with the 2PCF signals, with the strongest correlations at $\sim0.2R_{200}$. For larger galactocentric radii, the correlations become gradually weaker. $z_{3/4}$ shows the strongest correlations with the 2PCF. However, the correlations show a large scatter.

By looking at individual cases, we can see that for galaxies with similar 2PCF signals, their mass accretion histories can be either similar or different, but we observe that for the galaxies with later formation times, the 2PCFs are more likely to be higher in amplitude. We also see cases when the 2PCF signals are very different, the mass accretion histories are not significantly different. We show an extreme example of one galaxy with earlier formation time has significantly higher 2PCF signals than the other. These all suggest that although 2PCF signals correlate with the mass accretion history, the scatters are large.  

We investigate the 2PCF signals in velocity space, and find similar trends as the real space 2PCF. The velocity space 2PCF signals show slightly stronger correlations with the formation times beyond $\sim0.3-0.35R_{200}$. On the other hand, the real space 2PCF signals show better correlations within $0.3R_{200}$. This indicates that the velocity space clustering are slightly better preserved at larger galactocentric radii. However, the velocity space 2PCF signals carry on average less information than the real space 2PCF. Therefore, the velocity space 2PCF will not significantly increase the accuracy determining the galaxy assembly histories.

The real and velocity space 2PCF signals of halo stars also correlate with the assembly histories of their host dark matter halos, though still with large scatters. Within $0.3R_{200}$, the real space 2PCF shows stronger correlations with the galaxy formation histories than with the halo formation histories, while the velocity space 2PCFs do not show large differences in their correlations with the galaxy or halo formation times. It indicates that the velocity space 2PCF signals have similar performance upon predicting the galaxy or halo formation histories. 

We emphasize that throughout our analysis, observational errors are not incorporated. Real observational errors may further act to weaken the correlations between the 2PCF signals and the mass assembly histories. 

We conclude that it is difficult to use 2PCF statistics alone to precisely predict the formation times or assembly histories of galaxies or their host dark matter halos. However, higher order statistics such as three point correlation functions might further help to capture the non-linear process, and improve the precisions in predicting galaxy assembly histories.

\acknowledgments
This work is supported by NSFC (12022307, 12273021, 11973032, 11890691), the CSST Project grant No. 
CMS-CSST-2021-A02 and CMS-CSST-2021-A03, 111 project (No. B20019) and Shanghai Natural Science Foundation 
(No. 19ZR1466800). We thank the sponsorship from Yangyang Development Fund. 
The computation of this work is carried out on the \textsc{Gravity} supercomputer 
at the Department of Astronomy, Shanghai Jiao Tong University (DoA, SJTU). We thank useful technical support by the supercomputer administrator, Yihe Wang. W-W is 
very grateful for useful discussions with Andrew P. Cooper and Ling Zhu. 
Y-Z achieved this work as an under-graduate student in DoA, SJTU. 
This paper is initiated by the Participation in Research Program (PRP No. 41) at SJTU.

%







\bibliography{master}{}
\bibliographystyle{aasjournal}

\clearpage



\end{document}